\documentclass[]{aa}

\usepackage{txfonts}
\usepackage{graphicx}

\begin{document}

\title{A Corona Australis cloud filament seen in NIR scattered light}
\subtitle{II. Comparison with sub-millimeter data}

\author{
Mika Juvela \inst{1} \and
Veli-Matti Pelkonen \inst{1} \and
Sebastian Porceddu \inst{1}
}

\institute{
Helsinki University Observatory, FI-00014, University of Helsinki, Finland
}

\date{Received 1 January 2005 / Accepted 2 January 2005}

\abstract
{
Dust emission and scattering provide tools for the study of interstellar
clouds. Detailed knowledge of grain properties is essential for correct
analysis of these observations. However, dust properties are suspected to
change from diffuse to dense regions. The changes could affect our
interpretation of observations from large scales (e.g., clump mass spectra)
down to the structure of individual cores.
}
{
We study a northern part of the Corona Australis molecular cloud that consists
of a filament and a dense sub-millimetre core inside the filament. Our aim is
to measure dust temperature and sub-mm emissivity within the region. We also
look for confirmation that near-infrared (NIR) surface brightness can be used
to study the structure of even very dense clouds.
}
{
We extend our previous NIR mapping south of the filament. The dust colour
temperatures are estimated using Spitzer 160\,$\mu$m and APEX/Laboca
870\,$\mu$m maps. The column densities derived based on the reddening of
background stars, NIR surface brightness, and thermal sub-mm dust emission are
compared. A three dimensional toy model of the filament is used to study the
effect of anisotropic illumination on near-infrared surface brightness and the
reliability of dust temperature determination.
}
{ 
Relative to visual extinction, the estimated emissivity at 870\,$\mu$m is 
$\kappa_{\rm 870}= (1.3 \pm 0.4) \times 10^{-5}$\,${\rm mag^{-1}}$.
This is similar to the values found in diffuse medium. A significant increase
in the sub-millimetre emissivity seems to be excluded. In spite of saturation,
NIR surface brightness was able to accurately pinpoint, and better than
measurements of the colour excesses of background stars, the exact location of
the column density maximum. Both near- and far-infrared data show that the
intensity of the radiation field is higher south of the filament.
}
{}


\maketitle

\section{Introduction}

The column densities of dense molecular clouds can be estimated in several
ways. Because the dust can usually assumed to be well mixed with the gas, dust
can be used as a tracer of cloud structure. The available tools are based on
dust extinction, thermal dust emission, and light scattering. 

Extinction can be traced with star counts. However, in the NIR regime and for
dense clouds it is more common to make use of the reddening that takes place
when the light from background stars travels through a dust cloud. The method
is useful particularly in NIR because, at those wavelengths, the variation in
the intrinsic colours of background stars is small, the NIR extinction curve
is relatively insensitive to changes in the dust populations, and the lower
optical depths allow studies of clouds with visual extinctions up to $\sim
20^{\rm m}$. The reliability can be further increased by using several NIR
bands, the combination of J-, H-, and K-bands being the most common one (e.g.,
Lombardi \& Alves \cite{Lombardi2001}, Lombardi \cite{Lombardi2009}).

A complementary tool is provided by thermal dust emission at far-IR and
millimetre wavelengths. The emission is optically thin and, therefore, capable
of probing higher densities than what are accessible by observations of
background stars. The conversion between intensity and column density requires
knowledge of dust temperature and dust emissivity. The colour temperature can
be estimated from observations of two or more wavelengths. However, there are
always temperature variations along the line-of-sight and, because warm dust
produces much stronger signal than cold dust, the derived colour temperature
can be significantly higher than the actual temperature of the bulk of dust.
The problem is acerbated in pre-stellar cores where FIR observations can
completely ignore the cold core where the temperature can be below 10\,K or,
at the very centre, even close to 6\,K (Crapsi et al. \cite{Crapsi2007}; Harju
et al. \cite{Harju2008}). In geometrically simple cases, modelling can be used
to estimate the temperature variations and to correct the column density
estimates accordingly (e.g., Stamatellos et al.~\cite{Stamatellos2007})

Dust emissivity and the spectral index appear to vary from region to region
(Cambr\'esy et al. \cite{Cambresy2001}, del Burgo et al. \cite{delBurgo2003},
Dupac et al. \cite{Dupac2003}; Kramer et al. \cite{Kramer2003}, Stepnik et al.
\cite{Stepnik2003}; Lehtinen et al. \cite{Lehtinen2004}, \cite{Lehtinen2007};
Ridderstad et al. \cite{Ridderstad2006}). The variations are attributed to
either grain growth (Ossenkopf \& Henning \cite{ossenkopf1994}, Krugel \&
Siebenmorgen \cite{Krugel1994}) or physical changes in the grain material
itself (e.g., Mennella et al. \cite{Mennella1998}, Boudet et al.
\cite{Boudet2005}). Like temperature variations, changes in grain emissivity
should be correlated with density and, therefore, might cause systematic
errors in, for example, core mass spectra and the estimated structure of
pre-stellar cores.

With deep NIR observations one can detect the surface brightness that is
caused when dust grains scatter photons originating in the general
interstellar radiation field. Both numerical modelling and first observational
studies suggest that also the phenomenon can be used for quantitative study of
cloud structure (Padoan et al. \cite{Padoan2006}; Juvela et al.
\cite{Juvela2006}, \cite{Juvela2008}; Nakajima et al. \cite{Nakajima2008}).
The method works best when visual extinction is above a few magnitudes but
below $\sim 20^{\rm m}$. The lower limit is set by what is observationally
feasible and the upper limit by the saturation of the NIR signal at high
optical depths. In the absence of strong local radiation sources, the
reliability of the method appears to be comparable to the other methods listed
above. However, because the surface brightness is observed in NIR, the
obtained resolution is higher, potentially even of the order of one arc
second.

We continue the study of the northern filament of the Corona Australis cloud.
In previous paper Juvela et al. (\cite{Juvela2008}), we carried out comparison
between two column density estimators, one based on observations of background
stars and one on the scattered light. In this paper, we present new
observations that extend the previous NIR mapping south of the Corona
Australis filament. We have also mapped the region at 870\,$\mu$m using the
Laboca instrument on the APEX telescope. The morphology of the sub-millimeter
map is compared with FIR and NIR observations. With the aid of Spitzer
160\,$\mu$m data, we estimate the dust emissivity at 870\,$\mu$m. A
three-dimensional radiative transfer model is used to study the effects of
asymmetric illumination and the differences that may exist between the colour
temperature estimates and the true dust temperature.

\section{Observations} \label{sect:observations}

\subsection{Near-infrared observations}

\begin{figure} 
\resizebox{\hsize}{!}{\includegraphics{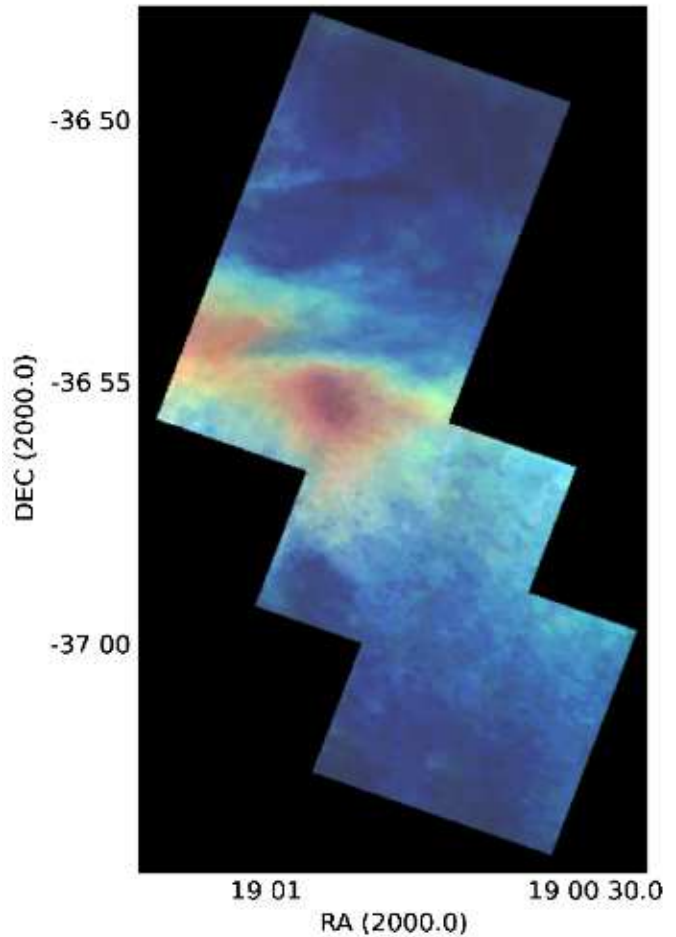}}
\caption{
A composite of the near-infrared observations of the Corona Australis
filament. The J-, H-, and Ks-bands are coded, respectively, in blue, green,
and red. The filament runs horizontally across the image and, because of the
stronger saturation of the J- and H-band surface brightness, shows up as a red
region.
} 
\label{fig:JHK} 
\end{figure}

A composite image of the near-infrared J-, H-, and Ks-band data is shown in
Fig.~\ref{fig:JHK}. The northern part of the field, some $4\arcmin \times
8\arcmin$ in size, was observed in August 2006 using the SOFI instrument on
the NTT telescope. Those data cover the densest part of the filament and, in
the north, extend to an area where the estimated visual extinction is $\sim
1^{\rm m}$. These observations were described in Juvela et
al.~\cite{Juvela2008}.

Further SOFI observations were made in June 2007. The new data consist of two,
partially overlapping $\sim 4\arcmin \times 4\arcmin$ fields. These extend the
original mapping south of the centre of the filament. As previously, the
observations were carried out as ON-OFF measurements so that the faint surface
brightness could be recovered. Three OFF fields were selected based on IRAS
images and 2MASS extinction maps from regions of low cirrus contamination. In
order to minimize the effect of sky variations, observations of ON- and
OFF-fields were interleaved in a sequence of OFF -- ON-2 -- ON-1 . The
OFF-field was varied between observation blocks to average out any faint
background gradients that might exist in the OFF-fields. Each observed frame
was an average of six 5 second exposures, corresponding to an integration time
of 30 seconds. After each sequence, the telescope returned to the
OFF-position, and, at the end of the observation block, took six 5 second
exposures of the OFF-field. The total integration times for each ON-field was
26 minutes in J, 37.5 minutes in H, and 97.5 minutes in Ks band (see
Table~\ref{table:fields}). The observations were calibrated to 2MASS scale by
using photometry of ten selected stars in each individual frame. Their average
was used to correct the observed fluxes to above the atmosphere. The two
ON-fields were mosaiced together using the overlapping area.  To have the same
signal in the overlapping area, this required an additional small correction
to ON-1 frame.
The overlapping area between the 2006 and 2007 observations, slightly more
than $1\arcmin \times 1\arcmin$ in size, was used for the final adjustment of
the new surface brightness measurements.

\begin{table}
\caption{Positions and total integration times of the observed ON and OFF fields.
The last column gives list for the OFF fields the extinction estimates of
Schlegel, Finkbeiner \&  Davis (\cite{SFD}).}
\begin{tabular}{llllll}
Field & centre position & $t$(J) & $t$(H) & $t$(Ks)
& $A_{\rm V}$ \\
&    (J2000)      &  (min)           & (min)            & (min) &  (mag) \\
\hline
ON-1    &  $19^{\rm h}0^{\rm m}42^{\rm s}$, $-37\degr 0\arcmin 54\arcsec$
& 26  & 37.5  & 97.5 &   -- \\
ON-2    &  $19^{\rm h}0^{\rm m}47^{\rm s}$, $-36\degr 57\arcmin 54\arcsec$
& 25.5  & 37.5  & 97.5 &   -- \\
OFF-1   &  $18^{\rm h}58^{\rm m}14^{\rm s}$, $-36\degr 50\arcmin 11\arcsec$
& 11  & 16  & 35  &   0.45 \\
OFF-2   &  $18^{\rm h}58^{\rm m}15^{\rm s}$, $-36\degr 34\arcmin 43\arcsec$
& 12  & 16  & 35  &  0.35 \\
OFF-3   &  $18^{\rm h}56^{\rm m}59^{\rm s}$, $-36\degr 53\arcmin 5\arcsec$
& 5.5  & 8  & 35  &  0.33 \\
\hline
\end{tabular}
\label{table:fields}
\end{table}

\subsection{Sub-millimeter and far-infrared data}

The Corona Australis filament was mapped with APEX/Laboca at 870\,$\mu$m. The
first observations were carried out in November 2007. However, the weather
conditions were not optimal and, additionally, some detector scans were
affected by the bright R Corona Australis region that is located some
15$\arcmin$ west of our NIR field. In the following, we rely mostly on data
gathered in October 2008. Observations consisted of scans $\sim$30$\arcmin$ in
length. These were made mainly in the North-South direction, i.e.,
perpendicular to the filament. To improve sky noise suppression, the position
angle of the scans was varied by up to $\sim$30 degrees. The precipitable
water vapour content of the atmosphere was $\sim$1.2\,mm. The observations
were reduced using the Boa program, version 1.1\footnote{see
http://www.astro.uni-bonn.de/boawiki}. The atmospheric opacities derived from
skydips using Boa are known to be underestimated. Following instructions on
the official Laboca web-site, the opacities were scaled up by a factor 1.3.

\begin{figure} 
\resizebox{\hsize}{!}{\includegraphics{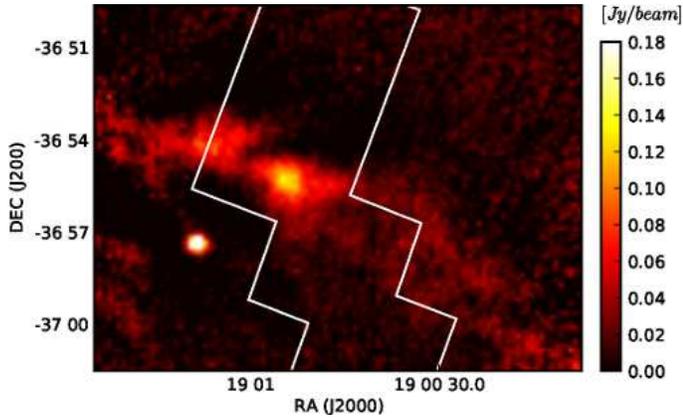}}
\caption{
The 870\,$\mu$m Laboca map. The white line denotes the outline of the NIR map
(see Fig.~\ref{fig:JHK}). The bright point source T Tauri star S Cr\,A (Joy
\cite{Joy1945}; Carmona et al.~\cite{Carmona2007}). 
} 
\label{fig:laboca} 
\end{figure}

During observations, Uranus, Neptune, and, as a secondary calibrator,
IRAS16293, were observed at regular intervals. The final maps contain one
bright millimeter point source, S Cr\,A, which gives possibility for another
check of calibration. The source is a T Tauri star that corresponds to the
submillimeter source MMS-7 observed by Nutter et al. (\cite{Nutter2005}) with
the SCUBA instrument. At 850\,$\mu$m, their estimate of the source flux was
0.70$\pm$0.02\,Jy/beam. Taking into account the difference in the beam size
and in the wavelength, our estimate is higher by slightly less than 8\%. The
870\,$\mu$m map is shown in Fig.~\ref{fig:laboca}.

The area has been mapped in Spitzer guaranteed time programs, at 3.6, 4.5,
5.8, and 8.0 $\mu$m with the IRAC instrument and at 24, 70, and 160\,$\mu$m
with the MIPS instrument. In this paper, we use only the longest wavelength,
160\,$\mu$m, to derive temperatures of the large dust grains that are
responsible for the sub-mm signal. The shorter Spitzer wavelengths probe small
grain and PAH populations. These should be included in a comprehensive model
of the cloud but are not used here, because they are largely independent of
the large grain population and, compared with the Laboca data, trace outer
cloud layers.

\section{Results}  \label{sect:results}

\subsection{NIR observations} \label{sect:NIR}

\begin{figure*} 
\resizebox{\hsize}{!}{\includegraphics{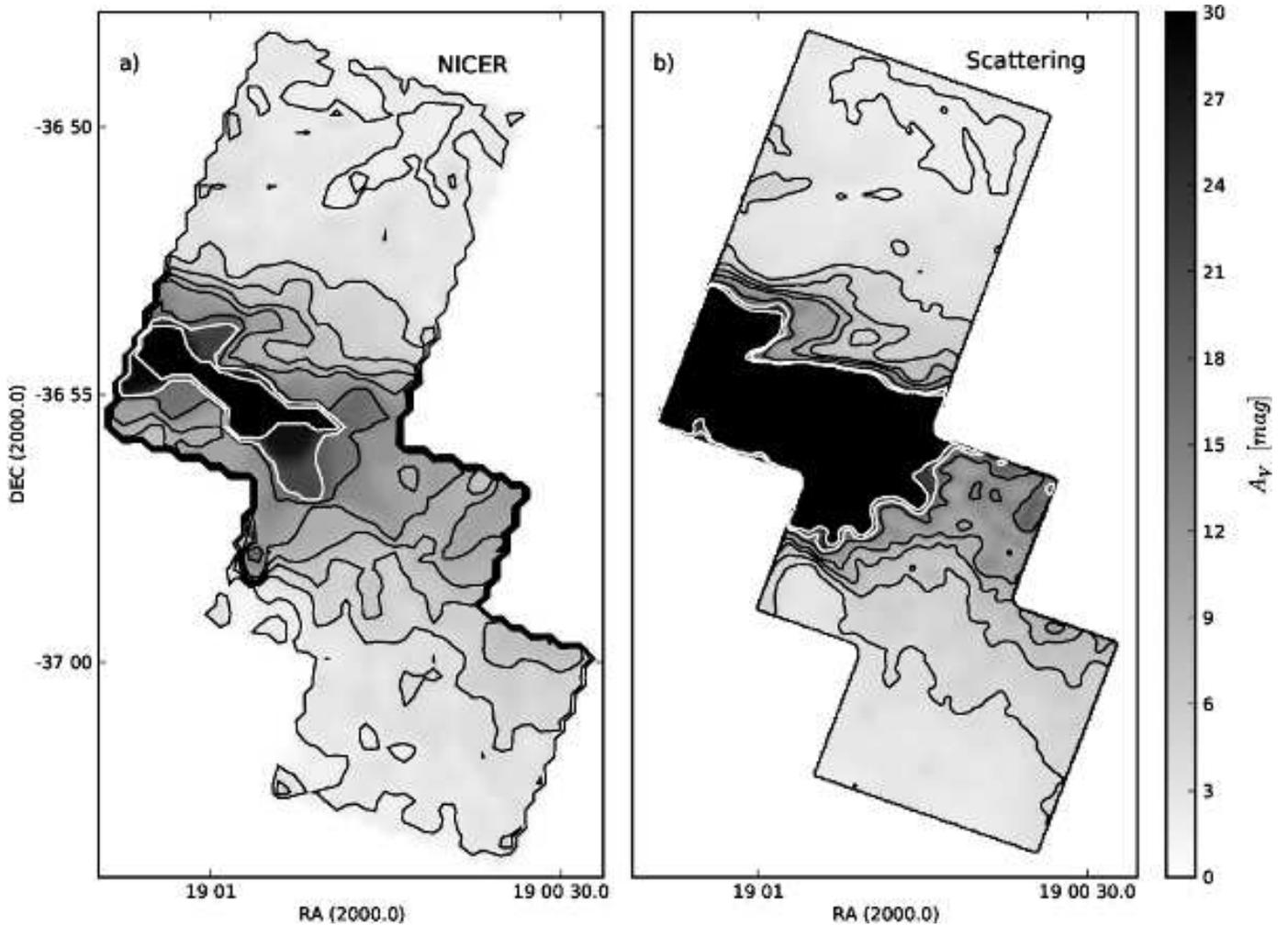}}
\caption{
Extinction maps obtained using the background stars (the NICER method; frame
$a$) and observations of scattered light (frame $b$). The black contours are
drawn at 2, 4, 6, 8, 10, and 15 and the white contours at 20 and 30 magnitudes
of visual extinction. The resolution is 20$\arcsec$ for the NICER map and
10$\arcsec$ in for the extinction maps derived from surface brightness
measurements.
} 
\label{fig:extinction} 
\end{figure*}

\begin{figure} 
\resizebox{\hsize}{!}{\includegraphics{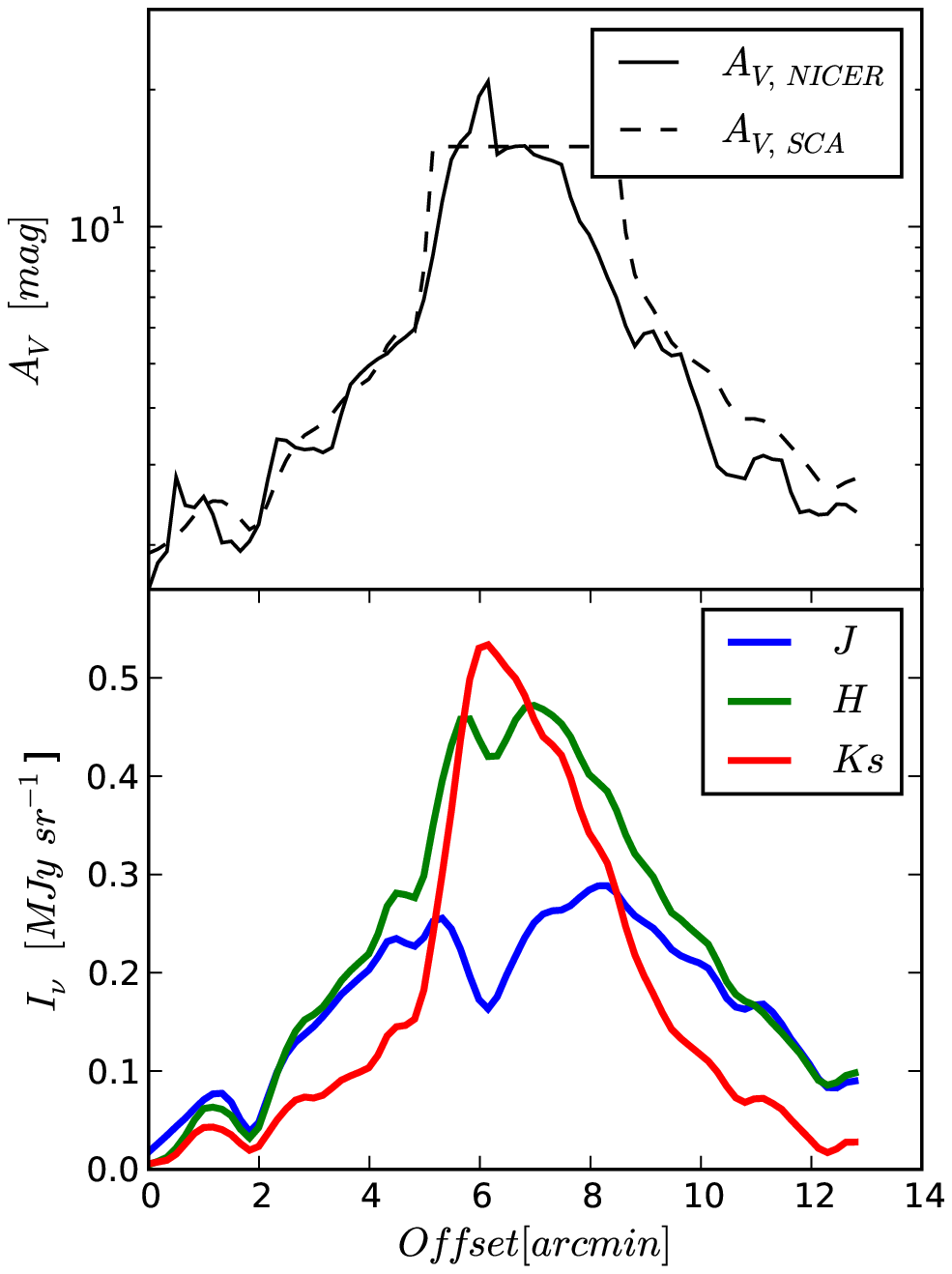}}
\caption{
{\em Upper frame:} Comparison of $A_{\rm V}$ profiles of the cloud filament
derived using the NICER method and using the scattered surface brightness.
For surface brightness, values are truncated at $A_{\rm V}=15^{\rm m}$ because
of the saturation of the scattered signal.
{\em Lower frame:} The profile of the filament in three NIR bands.
The cut is from -36$\degr$49$\arcmin$12$\arcsec$ in the north (offset zero) to
-37$\degr$2$\arcmin$21$\arcsec$ in the south. Values are averaged over the
interval from 19$^{\rm h}$0$^{\rm m}$43.8$\arcsec$ to 19$^{\rm h}$0$^{\rm
m}$50.4$\arcsec$.
} 
\label{fig:profiles} 
\end{figure}

In paper I (Juvela et al. \cite{Juvela2008}), good agreement was found between
$A_{\rm V}$ estimates obtained from the reddening of background stars and from
the surface brightness of scattered light. With the new observations, the
comparison can be extended to the southern side of the filament. For the
background stars, $A_{\rm V}$ estimates were derived using the NICER method
(Lombdardi \& Alves. \cite{Lombardi2001}). The surface brightness was
converted to $A_{\rm V}$ using the same method and exactly the same parameters
as in paper I. The extinction maps are shown in Fig.~\ref{fig:extinction}
while Figure~\ref{fig:profiles} shows North-South profiles of the filament in
$A_{\rm V}$ and scattered surface brightness.

As concluded in paper I, on the northern side (offsets 0--5$\arcmin$) the
methods give very similar extinction values. Within the filament the surface
brightness becomes gradually saturated (see Fig.~\ref{fig:JHK}) and the
uncertainty of the $A_{\rm V}$ estimates increases. Therefore, for the surface
brightness data, predictions are shown only up to $A_{\rm V}=15^{\rm m}$.
Within the filament, the density of background stars decreases. This leads to
bias also in the NICER estimates, as discussed in Paper I. Furthermore, in the
region where the Ks-band surface brightness is completely saturated also no
background stars are seen. Therefore, the true peak extinction can be higher
than shown by the NICER values.

On the southern side, the surface brightness predicts $\sim$20-30\% higher
extinctions than the colour excesses of the background stars. Compared to the
NICER extinction profile, especially in the J- and H-bands, the surface
brightness is higher on the southern side of the filament (see
Fig.~\ref{fig:profiles}, lower frame). This suggests that the intensity of the
radiation field would be higher on the southern side of the filament.
The conversion between scattered intensity and $A_{\rm V}$ depends on 
assumptions of dust properties and the illuminating radiation field. In Paper
I, the parameters were fixed using data north of the filament. Therefore,
assuming identical dust properties, the difference in the $A_{\rm V}$
estimates would indicate a similar percentual difference in the intensity of
the NIR radiation field. In the K-band the asymmetry is smaller than in the
other two bands. This suggests a change in the spectrum of the illuminating
radiation although the NIR colour ratio might also be affected by
observational uncertainties.

Both the new NIR observations and the existing far-infrared data (see
Sect.~\ref{sect:submm}) indicate that the intensity of the radiation field
increases towards south. In paper I an opposite conclusion was reached on the
basis of NIR colours close to the dip in the Ks-band intensity. The intensity
ratio J/Ks was found to be higher in the northern part of the dip which,
because the cloud is more transparent for Ks-band radiation, would seem to be
consistent with most of the illumination coming from that direction. However,
the situation may be more complicated and the outcome depends on whether the
observed local minimum of K-band surface brightness exactly coincides with the
extinction maximum or is shifted away from the direction of stronger
illumination. We return to this question in Sect.~\ref{sect:RT} where we
examine the situation with the help of a radiative transfer model.

\subsection{Comparison of NIR, far-infrared, and sub-mm emission}
\label{sect:submm}

\begin{figure} 
\resizebox{\hsize}{!}{\includegraphics{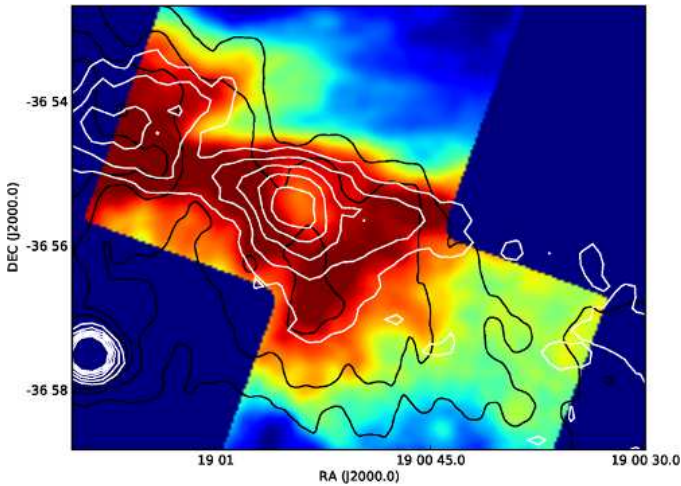}}
\caption{
The Ks-band surface brightness (colour image) with contours of 870$\mu$m
(white lines) and 160\,$\mu$m (black lines) surface brightness. For
870\,$\mu$m, the contours are from 0.02 to 0.12\,mJy/beam, at intervals of
0.02\,mJy/beam. The 160\,$\mu$m contours are drawn at 140, 160, 180, and
200\,MJy\,sr$^{-1}$.
} 
\label{fig:contour} 
\end{figure}

Figure~\ref{fig:contour} compares the morphology of Cr\,A filament as seen at
different wavelengths. The colour image shows the Ks-band surface brightness
that should be well correlated with the true column density up to $A_{\rm
V}\sim 10^{\rm m}$ but which, within the filament, becomes saturated and
exhibits a dip at the centre of the filament. In addition to the NIR data, the
figure shows Spitzer 160\,$\mu$ observations and our Laboca observations at
870\,$\mu$m. 

The sub-mm dust emission shows very good correlation with the morphology of
NIR emission. At the location of 870\,$\mu$m maximum, the NIR signal is
saturated but, based on the ratio of J, H, and Ks intensities, this was
already concluded to be the position of the highest column density (Juvela et
al.~\cite{Juvela2008}). In other words, the dip in the Ks-band image
pinpointed the location where the sub-mm emission was going to peak. 

In Sect.~\ref{sect:NIR} we concluded that the radiation field is stronger on
the southern side of the filament. In Fig.~\ref{fig:contour} the 870\,$\mu$m
emission is displaced by some 10$\arcsec$ south-east of the Ks-band dip. This
is consistent with the idea of asymmetric illumination: if the dip in NIR
surface brightness is shifted away from the direction of stronger radiation,
the sub-mm peak should similarly be shifted slightly towards the stronger
radiation. In Fig.~\ref{fig:contour} the asymmetry points towards the nearby T
Tauri star S Cr\,A as the cause of these shifts.  However, at 160\,$\mu$m the
emission is very extended on the souther side of the filament. This reflects
also the mass distribution that is asymmetric with respect to the densest part
of the filament. In the 160\,$\mu$m emission, the peak is directly south of
the sub-mm peak, not towards the star S Cr\,A. 
At shorter wavelengths, the 100\,$\mu$m surface brightness rises from
$\sim$20\,MJy\,sr$^{-1}$ in the northern part of our NIR field to
41\,MJy\,sr$^{-1}$ in the filament. However, south of the dense filament, the
surface brightness stays above 35\,MJy\,sr$^{-1}$. In the Spitzer 70\,$\mu$m
maps, the surface brightness rises as one moves southwards across the
filament, without clear correlation with the position of S Cr\,A.

\subsection{Dust colour temperature in the Corona Australis filament}
\label{sect:temperature}

\begin{figure} 
\resizebox{\hsize}{!}{\includegraphics{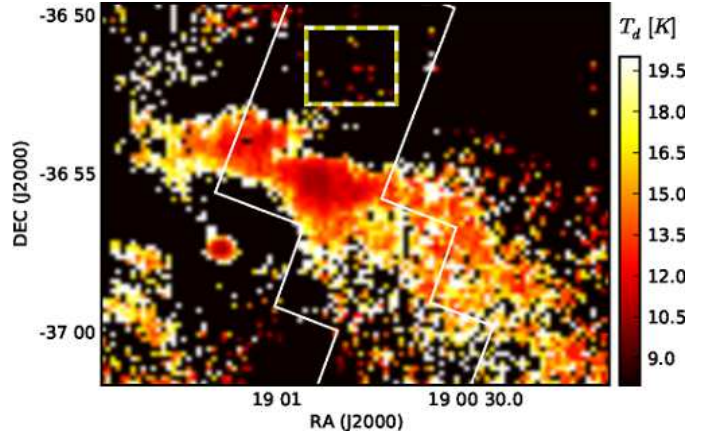}}
\caption{
The 870\,$\mu$m Laboca map. The white line denotes the outline of the NIR maps
(see Fig.~\ref{fig:JHK}). The bright point source T Tauri star S Cr\,A (Joy
\cite{Joy1945}). 
} 
\label{fig:temperature} 
\end{figure}

The colour temperature of large grains was estimated using the 160\,$\mu$m
Spitzer MIPS map and our APEX/Laboca 870\,$\mu$m observations. The values are
based on the assumption of a single gray body spectrum, $B_{\nu}(T) \times
\nu^{\beta}$. Unfortunately, the value of the spectral index is uncertain
(e.g., Dupac et al. \cite{Dupac2003}) and, because dust properties may exhibit
variations that are correlated with density and temperature, $\beta$ may have
significant variations even within a single cloud. There is also indication
that the spectral index is wavelength dependent and that, compared to sub-mm
and millimetre wavelengths, the variation of $\beta$ is stronger in the FIR
regime. With the available data, we cannot study these variations in detail.
The colour temperature map obtained for a fixed value of $\beta=2.0$ is shown
in Fig.~\ref{fig:temperature}. The figure also indicates the area that was
used for the determination of the intensity zero level at the two wavelengths.

In the reduced 870\,$\mu$m map, only the densest filament is visible. On both
sides, the intensity first becomes slightly negative before again raising
towards north and south. This suggests that, in the data reduction, the
baseline subtraction was not perfect. However, the overshoot to negative
values is only $\sim$5\% of the peak intensity and, for the colour temperature
calculation, the zero level is determined from an area where the surface
brightness reaches minimum (see Fig.~\ref{fig:temperature}). Therefore, the
uncertainty in the baseline subtraction translates to an uncertainty in colour
temperature that is only $\sim$0.1\,K for the sub-mm peak but can be close to
one degree in areas of lower intensity where the colour temperature is close
to 15\,K. A related source of uncertainty results from the fact that, in the 
870\,$\mu$m map, some of the large scale structure may have been filtered out.

In the filament, the colour temperature is reduced to $T_{\rm C}=$10.7\,K. In
this value, the uncertain value of $\beta$ is a larger source of uncertainty
than the observational errors. Varying $\beta$ in the plausible range
from 2.5 to 1.5 would change the minimum colour temperature between 9.5 and
12.4\,K. 
A change $\Delta \beta$=0.5 corresponds to a change by a factor of 2.7 in the
relative dust opacity at 160\,$\mu$m and 870\,$\mu$m.
Because of temperature variations along the line-of-sight, the colour
temperature is likely to overestimate the mass-weighted true dust temperature.
This bias is examined further in Sect.~\ref{sect:RT}, in connection with a
radiative transfer model.

\subsection{Sub-mm dust emissivity} \label{sect:emissivity}

\begin{figure} 
\resizebox{\hsize}{!}{\includegraphics{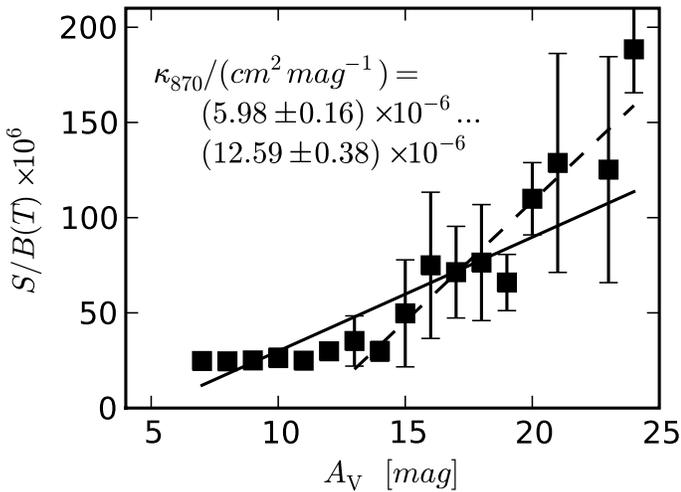}}
\caption{
Ratio of 870\,$\mu$m surface brightness and the Planck
function, $B$($T_{\rm C}$), as function of visual extinction.  Least square
fits are shown for all data with $I_{\rm nu}(870)>0.01$\,Jy/beam and for the
central part of the filament where $A_{\rm V}>13^{\rm m}$.
} 
\label{fig:emissivity} 
\end{figure}

Using the derived colour temperatures $T_{\rm C}$ and the surface brightness
$I_{\nu}$, the dust emissivity at 870\,$\mu$m, relative to visual extinction,
can be calculated differentially,
\begin{equation}
         \kappa_{870} 
         =  
         \frac{ \Delta I_{\nu}/B_{\nu}(T_{\rm C}) }{\Delta A_{\rm V}} \, \,
         \, \, 
         [{\rm mag}^{-1}].
\end{equation}
This assumes that dust emissivity remains constant over the whole area.
Figure~\ref{fig:emissivity}$a$ shows the correlation between 
$I_{\nu}/B_{\nu}(T_{\rm C}$ and visual extinction. 
To lower the noise of $A_{\rm V}$, we use weighted average of the maps shown
Fig.~\ref{fig:extinction}. At low $A_{\rm V}$, equal weight is given to the
two maps but, by $A_{\rm V}\sim 23^{\rm m}$, the values are based on NICER
method alone.
The slope of the linear fit gives a value $\kappa_{\rm 870}=0.6 \times
10^{-5}$\,${\rm mag^{-1}}$. However, the slope steepens towards the filament
and, for $A_{\rm V}>13^{\rm m}$, one obtains emissivity that is twice as high,
$\kappa_{\rm 870}=1.3 \times 10^{-5}$\,${\rm mag^{-1}}$.

Emissivity can also be estimated directly on the basis of absolute values. As
above, the values are low outside the sub-millimeter peak but increase to
$\sim 1.1 \times 10^{-5}$\,${\rm mag^{-1}}$ at the location of the surface
brightness maximum. In this case, the zero level of surface brightness values
was estimated north of the filament, around declination -36$\degr$53$\arcmin$.

There is a temptation to interpret the spatial variations as an indication of
increase of dust sub-mm emissivity towards the densest parts of filament.
However, the 870\,$\mu$m surface brightness is detected only close to the
actual sub-mm peak and the emissivities, obtained either from correlation or
absolute surface brightness, are uncertain outside this region. Furthermore,
in the 870\,$\mu$m map, some of the large scale structure may have been
filtered out. This could account for the lower value obtained given by the
absolute surface brightness values and could contribute to the decrease seen
outside the surface brightness maximum. Therefore, considering both the S/N
ratio and the possible filtering out of large scales, the values obtained at
the very centre of the filament should be the most reliable ones.  

The emissivities were calculated using the 160\,$\mu$m/870\,$\mu$m colour
temperature which may overestimate the true dust temperature if there are
strong temperature variations along the line-of-sight. In Sect.~\ref{sect:RT}
we will examine, whether this could lead to {\em underestimation} of
$\kappa_{\rm 870}$. In paper I, we concluded that, extinction within the
filament may be underestimated because of the steep extinction gradients. The
bias, which was estimated to be $\sim$25\% at $A_{\rm V}\sim 20^{\rm m}$,
would cause $\kappa_{870}$ to be {\em overestimated}, probably by a similar
factor.
We quote emissivities relative to visual extinction. In the NICER method, the
conversion to $A_{\rm V}$ was done assuming a standard extinction curve that
corresponds to $R_{\rm V}$=3.1. Although it is probable that $R_{\rm V}$
increases towards the dense filament, the effect on $\kappa/A_{\rm V}$ is 
only of the order of one percent, i.e., negligible compared with the other
sources of uncertainty.

Because of the various sources of uncertainty, we can only conclude that, at
the centre of the Cr\,A filament, the dust emissivity is likely to be at least
$\sim 1 \times 10^{-5}$\,${\rm mag^{-1}}$ but could easily be higher by
$\sim$50\%. Relying on the values obtained at the centre of the filament and
taking into account the various biases, we write our estimate as 
$\kappa_{\rm 870}= (1.3 \pm 0.4) \times 10^{-5}$\,${\rm mag^{-1}}$.
The quoted uncertainty is, of course, only approximative.

\section{Toy model of the filament} \label{sect:RT}

A toy model of the Cr\,A filament was used to examine two questions, the effect
of asymmetric illumination on the observed NIR surface brightness and the bias
of the dust colour temperature used in the derivation of sub-mm dust
emissivity.
The model consists of a cylinder, representing the large-scale filament, and a
denser spherical region in the centre, representing the observed sub-mm peak.
The density follows gaussian profile, both with respect to distance from the
axis of the cylinder and from the centre of the spherical component. The
extinction is $\sim$10$^{\rm m}$ across the cylinder and reaches $\sim
70^{\rm m}$ through the centre, i.e., through the combined density peak of the
cylinder and the sphere. The model was resampled onto a cartesian grid of
65$^3$ cells.

The dust properties are taken from Draine (\cite{Draine2003}). The scattering
was calculating using the tabulated phase function available on the web
\footnote{http://www.astro.princeton.edu/$\sim$draine/dust/}, assuming the
dust model corresponding to $R_{\rm V}=5.5$. For radiative transfer
calculations, we use our Monte Carlo programs (Juvela \&
Padoan~\cite{Juvela2003}; Juvela~\cite{Juvela2005}).

\subsection{Correlation of NIR surface brightness}

In Paper I the spatial variations in the NIR intensity ratios were examined and
several regions were identified that exhibited a particular combination of
surface brightnesses. Using the toy model, we re-examine the interpretation of
the regions $c-d$ and $f-g$ that were shown in Fig.~4 of that paper.

\begin{figure} 
\resizebox{\hsize}{!}{\includegraphics{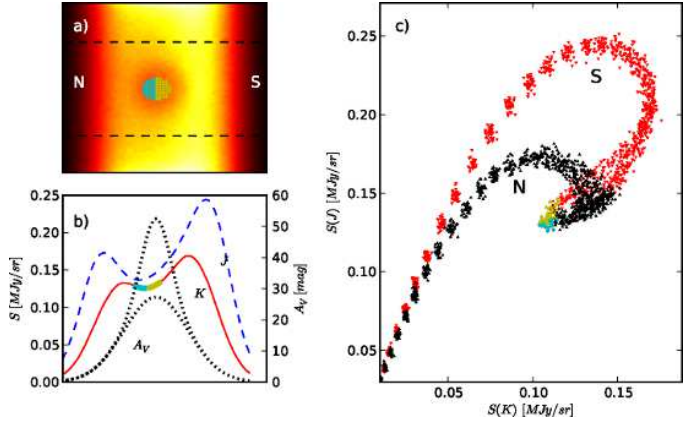}}
\caption{
NIR surface brightness in our toy model.
{\bf a.} Image of the K-band surface brightness. The cylindrical filaments
runs vertically across the image and an additional density peak is located at
the centre of the image. The illumination is stronger on the right hand side
of the filament.
{\bf b.} Surface brightness and visual extinction profiles of the filament
averaged over the area limited by the dashed lines drawn in frame {\em a.}.
The lower dotted line shows $A_{\rm V}$ cross section of the cylinder alone.
{\bf c.} Correlation of $J$- and $K$-band surface brightness for pixels
between dashed lines in frame {\em a}. 
Colours highlight pixels close to the $A_{\rm V}$ peak.  Yellow triangles (tip
pointing upwards) correspond to pixels on the side of higher illumination and
cyan triangles (tip pointing downwards) to pixels away from the stronger
illumination. In frames $a$ and $c$, the letters $N$ and $S$ refer to north
and south, following the hypothesis that the illumination of the Cr\,A
filament is stronger in the south.
} 
\label{fig:NIR_model} 
\end{figure}

Figure~\ref{fig:NIR_model} shows the correlation between J- and K-band surface
brightness in the toy model. The model cloud is illuminated by an isotropic
radiation field that corresponds to the NIR ISRF given in Lehtinen et al.
(\cite{Lehtinen1996}). Additionally, in the south (as marked in the figure) 
there is an additional radiation source with similar spectrum that raises the
total intensity by 30\%. This introduces asymmetry that is clearly visible as
northern and southern branches in the correlation plotted in
Fig.~\ref{fig:NIR_model}c.

Figure~\ref{fig:NIR_model}c can now be compared qualitatively with Fig.~4 of
paper I. The regions $c$ and $d$ are clearly consistent with asymmetric
illumination with stronger radiation in the south. More specifically, the
asymmetry suggests that the additional radiation is coming from south-east.

For regions $f$ and $g$ the correspondence with the toy model is less clear. 
In the model, the northern and southern branches (as marked in
Fig.~\ref{fig:NIR_model}c) cross each other before the location of peak
extinction is reached. Therefore, the intensity ratio J/K should be higher on
that side of the $A_{\rm V}$ peak where the radiation is stronger. If regions
$f$ and $g$ are interpreted as the tip of the two branches, they would
perfectly match a case where radiation comes mainly from the north rather than
from the south. Because other evidence clearly shows that this is not the
case, the final interpretation of the NIR colours at the centre of the
filament remains open. Smaller scale density structures (e.g., structures
extending in front of the main cloud) could provide one possible explanation.

In the model the southern radiation source extended 10$\degr$ from the
southern direction. Figure~\ref{fig:NIR_model} remains qualitative similar
even if the opening angle is as large as 45$\degr$. Similarly, a clear loop
that is formed in Fig.~\ref{fig:NIR_model}c) when the southern and northern
branches cross, persists even when the radiation source is moved $\pm 45
\degr$ from the south towards the front or back of the cloud. However, the 
the separation of the two branches is much smaller if the source is moved
towards the front of the cloud. If the source is moved towards the back of the
cloud, the peak surface brightness increases but the intensity at the centre
of the cloud decreases, both by up to $\sim$20\%.
These effects suggest that, with more detailed modelling of observations,
further constraints may still be derived for the radiation field.

\subsection{Colour temperatures from the toy model}

The determination of sub-mm emissivity can be affected by errors in the
estimated dust temperature. In particular, when there are temperature
variations along the line-of-sight, the observed colour temperature is known
to be biased towards higher temperatures (e.g.~Lehtinen et al.
\cite{Lehtinen2007}). We use the toy model to estimate the magnitude of this
bias. We use exactly the same model as above, i.e., including the additional
radiation source in the south that increases the total illumination by 30\%. 

\begin{figure} 
\resizebox{\hsize}{!}{\includegraphics{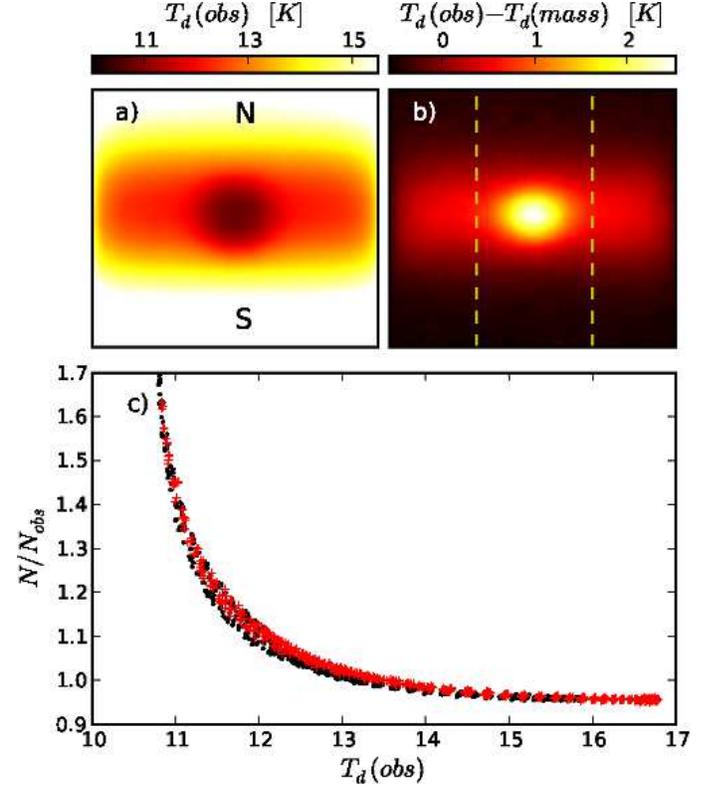}}
\caption{
Comparison of observed colour temperatures and the true dust temperature in
the toy model.
{\bf a.} The dust colour temperature based on 160\,$\mu$m and 870\,$\mu$m
surface brightness.
{\bf b.} The difference between the colour temperature and the mass
weighted average temperature along each line-of-sight.
{\bf c.} The ratio of true column density and the column density estimated
based on the observed colour temperature. The plot shows pixels that, in frame
$b$, fall between the dashed lines. Pixels south of the symmetry axis are
plotted as red plus signs.
} 
\label{fig:check_T} 
\end{figure}

Figure~\ref{fig:check_T}$a$ shows a map of colour temperatures that is
calculated from modelled surface brightness at 160\,$\mu$m and 870\,$\mu$m. A
value of $\beta=2.0$ is used for the emissivity index. This is very close to
the spectral index of the employed dust model, which is 1.96 between the two
frequencies. Frame $b$ shows the difference between the observed colour
temperature and the mass-weighted average temperature along each
line-of-sight. As expected, the colour temperature overestimates the true mean
temperature of dust grains. The error increases with opacity and is just over
2\,K at the centre of the cloud. 

Because the dust emission is optically thin, the ratio of the true column
density and the column density estimated on the basis of the colour
temperature can be calculated from
\begin{equation}
          N_{\rm true}/N_{\rm obs} =  
          \int_{\rm 0}^{s} n(s) B_{\nu}(T(s)) ds /
          [ B_{\nu}(T_{\rm C}) \int_{\rm 0}^{s} n(s) ds ].
\end{equation}
Here $T$ is dust temperature, $T_{\rm C}$ colour temperature, and $s$ distance
along the line-of-sight. Figure~\ref{fig:check_T}$c$ shows this ratio as
function of observed colour temperature. At the cloud centre, the true column
density is more than $\sim$50\% higher than what is estimated on the basis of 
colour temperature. This suggests that the value of $\kappa_{870}$ derived in
Sect~\ref{sect:emissivity} could be underestimated although not necessarily by
the same amount.

To estimate the effect on $\kappa_{870}$, we scaled the 870$\mu$m surface
brightness values upwards according to the relation of
Fig.~\ref{fig:check_T}$c$. If the Cr\,A filament behaves as our model, the
scaled surface brightness should correspond to a situation where all dust
along a line-of-sight has temperature equal to the colour temperature. 
Repeating the analysis of Sect.~\ref{sect:emissivity}, values of $\kappa_{\rm
870}$ increase less than 10\% in the fit to all data and $\sim$20\% in the fit
to data above $A_{\rm V}=13^{\rm m}$. The increase in the peak value of
Fig.~\ref{fig:emissivity} is similarly $\sim$20\%. While noticeable, the bias
caused by the use of colour temperatures may be compensated by the opposite
bias in $A_{\rm V}$ values. Therefore, the value derived in
Sect.~\ref{sect:emissivity}, 
$\kappa_{\rm 870}= (1.3 \pm 0.4) \times 10^{-5}$\,${\rm mag^{-1}}$,
remains our best estimates.

\section{Discussion} \label{sect:discussion}

The IRAS and Spitzer far-infrared observations demonstrate an asymmetry where
the intensity of the radiation field and/or the extent of diffuse material is
larger south of the Corona Australis filament. Our extinction measurements
show that the extinction is reduced below $A_{\rm V}\sim 3^{\rm m}$ within
5$\arcmin$ of the filament, both in the North and in the South. The column
densities are slightly larger in the south but NIR and far-infrared colours
indicate that the asymmetry is mainly caused by the illumination of the cloud.
Our toy model (Fig.~\ref{fig:NIR_model}) showed that, at large scales, the NIR
colour variations are consistent with this interpretation.  In the model, the
radiation field was increased by 30\% by adding one radiation source in the
south but this resulted in larger NIR colour variations than what was
observed. Either the asymmetry is, at NIR wavelengths, less strong or the
radiation comes from a much large solid angle or from a direction that
deviates significantly from the exact southern direction.


Stars S Cr\,A and HD 176269 lie a few arc minutes south-east of the filament.
HD 176269 is a pre-main sequence B9V star which, according to the HIPPARCOS
parallax, is at the distance of Corona Australis, $\sim 140$ pc. In Digital
Sky Survey Blue map, there is a reflection nebula around the star. The 2MASS
magnitudes of the star are 6.7, 6.8 and $6.7^{\rm m}$ for the J, H and Ks
bands, respectively. Assuming the intervening extinction is similar to the
extinction between Cr\,A and us and taking into account the projected distance
between the star and the outskirts of the filament, $\sim 3'$, the star could
produce an additional intensity that is $\sim 3.4 (2.1, 1.4) \times 10^{-18}
\; {\rm erg \, s^{-1} \, cm^{-2} \, Hz^{-1}}$ in J-band (H-band, Ks-band). This
would add a significant, bluer component to the ISRF.

The 70 $\rm \mu m$ surface brightness shows a clear surface brightness
gradient across the filament and brightening towards the south. This supports
the assumption that the additional illumination comes directly from the south,
rather than in more towards the front or back of the filament.  In the south,
the ratio of the J- and H-bands to the Ks-band decrease from east to west,
this also being consistent with bluer illumination from the star HD\,~176269.
S Cr\,A is located much closer to the dense part of the filament, but its
spectrum is redder and would not explain the different colour of scattered
light further south of the filament. The IRIS 100 $\rm \mu m$ map implies that
the influence of S Cr\,A to the heating of dust is restricted to its immediate
surroundings but it could have a contribution to the NIR flux impinging on
the filament.


In paper I we correlated NIR surface brightness with the extinction derived on
the basis of the colour excesses of background stars. We concluded that the
diffuse signal consists mainly of scattered light and it can be used for
quantitative study of cloud structure. Unless the method is calibrated using
colour excess data (see Juvela et al.~\cite{Juvela2008}; Nakajima et
al.~\cite{Nakajima2008}), the conversion to column density requires
assumptions of the radiation field and dust properties. The method should
remain reliable as long as the radiation field is roughly constant within the
examined region. In the Cr\,A field, this condition is clearly not fulfilled
because of the high opacity of the central filament and the asymmetry of the
illuminating radiation. In the south, assuming similar conditions as on the
northern side of the filament, the column density derived from NIR surface
brightness is overestimated by $\sim$20\% (see Fig.~\ref{fig:profiles}a).
However, the results from the surface brightness data can always be corrected
with the aid of colour excess measurements. This can be done at any scale down
to the resolution allowed by the spatial density of background stars. Compared
to the use of background stars only, this should result in a more accurate
column density measurements and, as far as allowed by the observational noise
of surface brightness data, extension to smaller spatial scales.

Our estimate for the 870\,$\mu$m emissivity was $\kappa_{\rm 870}= (1.3 \pm
0.4) \times 10^{-5}$ per magnitude of visual extinction. In the dust models
employed in our toy model, the 870\,$\mu$m dust extinction cross sections are
very similar, $1.3 \times 10^{-5}$\,${\rm mag^{-1}}$ for $R_{\rm V}=3.1$ and
$1.2 \times 10^{-5}$\,${\rm mag^{-1}}$ for $R_{\rm V}=5.5$. In this respect,
the models are consistent with the actual filament.

Sub-mm data obtained from PRONAOS experiment favours dust spectral index
values close to $\beta=2$ (e.g., Bernard et al.~\cite{Bernard1999}; Stepnik et
al. \cite{Stepnik2003}). The spectral index is, however, temperature dependent
and rises to $\sim$2 only when temperatures is below $\sim$20\,K (Dupac et
al.~\cite{Dupac2003}). There is some scatter in the values and, even at low
temperatures, the spectral index may vary within a wide range from $\sim$1.5
to $\sim$2.5. The determination of the spectral index is, of course, closely
related to the dust emissivity. As mentioned in Sect.~\ref{sect:temperature},
varying the value of $\beta$ in the range from 1.5 to 2.5 will cause the
minimum temperature to vary between 9.5 and 12.4\,K. In the value of $\kappa$,
this corresponds to an uncertainty of $\sim$30\%.
If sub-millimetre emissivity had shown significant increase, the true
spectral index $\beta$ would probably have been smaller than assumed. By
overestimating $\beta$, the dust temperature would have been underestimated
and the obtained value of $\kappa_{\rm 870}$ would have been even higher than
the true value.

The obtained value of $\kappa_{\rm 870}$ is similar to the values that have
been obtained in low density regions. Using COBE data, Boulanger et al.
(\cite{Boulanger1996}) derived effective dust cross section for dust in high
latitude diffuse medium,
\begin{equation}
  \sigma_{\rm H} = 10^{-25} (\lambda/250\,\mu m)^{-2} \, {\rm cm}^{-2}.
\end{equation}
Using a conversion factor $N({\rm H}_2)/A_{\rm V} = 0.94 \times
10^{21}$\,cm$^{-2}$\,mag$^{-1}$ (Bohlin et al. \cite{Bohlin78}), equally valid
for diffuse regions, this corresponds to $\sigma(870\,\mu{\rm m}) =1.55 \times
10^{-5}$\,mag$^{-1}$.

In the literature, higher emissivities are often reported for dense and cold
clouds. Stepnik et al. (\cite{Stepnik2003}) studied a filament in Taurus.
According to star counts, the extinction was at least $A_{\rm V}\sim
4^{\rm m}$ at the resolution of $3.5\arcmin$. However, using 2MASS H-K
colours, Padoan et al. (\cite{Padoan2002}) estimated the peak extinction to be
$\sim$12$^{\rm m}$.
Based on PRONAOS observations, Stepnik et al. concluded that the sub-mm
emissivity of big grains had increased by a factor of $\sim$3.4 within the
filament. At shorter wavelengths, Lehtinen et al. (\cite{Lehtinen2007})
reported a similar trend in the cloud L1642. The ratio of 200\,$\mu$m optical
depth and visual extinction increased by a factor of four when moving from
regions with $T_{\rm dust}\sim 20$\,K to dense regions with $T_{\rm dust}\sim
12$\,K. Kramer et al. (\cite{Kramer2003}) estimated variations of 850\,$\mu$m
emissivity in dark cloud IC~5146. The emissivity was found to change from
$\kappa_{850}/\kappa_{V} \sim 1.3\times 10^{-5}$ in warm regions to $\sim 5
\times 10^{-5}$ in colder regions with $T\sim12$\,K.
On the other hand, in their study of another Taurus filament, Nutter et al.
(\cite{Nutter2008}) found no such strong variations. The observations,
including Spitzer 70\,$\mu$m and 160\,$\mu$m and SCUBA 450\,$\mu$m and
850\,$\mu$m data, could all be modelled accurately without resorting to any
spatial variations of dust properties. The filament was similar to our Cr\,A
field and had higher extinction than the region studied by Stepnik et
al.~\cite{Stepnik2003}.

In our 870\,$\mu$m observations of the Cr\,A filament, part of large scale
structures may have been filtered out and this could lead to underestimation
of dust emissivity. However, this should not have a large effect at the short
spatial scales used in Sect.~\ref{sect:emissivity}. 
Our observations suggest that the emissivity in the Corona Australis filament
is similar to the values of diffuse medium. In spite of several sources of
uncertainty, it is very unlikely that the emissivity could be even twice as
high. 
This is not necessarily in strong contradiction with the results of Stepnik et
al. (\cite{Stepnik2003}) because our $\kappa_{870}$ is an effective value for
the whole line-of-sight. In the case of Stepnik et al., the emissivity of
large grains was increased by a factor 3.4 only at the centre of their
radiative transfer model. Although the observed value of $\kappa$ should be
dominated by this densest region, the effective value for the whole
line-of-sight should still be somewhat lower.
However, the studied CrA filament appears to be clearly different from the
dense cores of IC~5146. In the study of Kramer et al. (\cite{Kramer2003}),
$\kappa_{850}/\kappa_{V}$ was above $2\times 10^{-5}$ everywhere the dust
colour temperature (based on 450$\mu$m and 850\,$\mu$m observations) was below
14\,K.
These results indicate that the changes in sub-mm emissivity are not a
universal function of density or temperature.

In a future paper, we will construct a three-dimensional radiative transfer
model for the Cr\,A filament. Varying the density distribution, radiation
field, and dust properties, we will seek for a solution that would be
consistent with observations from NIR to sub-mm.

\section{Conclusions} \label{sect:conclusions}

We have complemented our previous NIR observations of the northern filament of
the Corona Australis molecular cloud. The dust extinction and NIR scattering
are now mapped both north and south of the filament. The observations,
together with existing far-infrared data, show that the illuminating radiation
field becomes stronger in the south.

The area was mapped at 870\,$\mu$m using the APEX Laboca instrument. With the
help of Spitzer 160\,$\mu$m maps, we estimate dust colour temperatures which,
within the central filament, decreases down to $T_{\rm C}\sim 10.5$\,K. On the
basis of the colour temperatures, the dust emissivity at 870\,$\mu$m is
estimated to be 
$\kappa_{\rm 870}= (1.3 \pm 0.4) \times 10^{-5}$\,${\rm mag^{-1}}$
as a value relative to $A_{\rm V}$. In spite of the high visual extinction,
the estimate is close to the values found in diffuse regions.

A three-dimensional toy model of the filament was presented.  The model
demonstrates that the main NIR colour variations described in paper I can be
explained as a result of anisotropic illumination.  The differences between
observed colour temperatures and mass-averaged true temperatures of large dust
grains were examined using the same model. The use of colour temperatures will
lead to underestimation of $\kappa_{\rm 870}$ values but the effect is no more
than $\sim$25\%. In the case of the Corona Australis filament, this could be
compensated by an opposite bias in the $A_{\rm V}$ estimates.

\acknowledgements
We thank the referee, Laurent Cambr\'esy, for helpful comments. M.J. and
V.-M.P. acknowledge the support of the Academy of Finland Grants no. 206049,
115056, and 107701.

\end{document}